\documentclass[a4paper,aps,prl,twocolumn,floats]{revtex4}
\usepackage{graphicx,dcolumn,bm}
\usepackage{amsmath,amssymb,amsfonts}
\usepackage[dvips]{hyperref}
\begin{document}

\title{\Large Asymmetric dark matter and the Sun}
\author{Mads T. Frandsen}
\author{Subir Sarkar}
\affiliation{Rudolf Peierls Centre for Theoretical Physics,
University of Oxford, 1 Keble Road,  Oxford OX1 3NP, UK}
\begin{abstract}
  Cold dark matter particles with an intrinsic matter-antimatter
  asymmetry do not annihilate after gravitational capture by the Sun
  and can affect its interior structure. The rate of capture is
  exponentially enhanced when such particles have self-interactions of
  the right order to explain structure formation on galactic scales. A
  `dark baryon' of mass 5 GeV is a natural candidate and has the
  required relic abundance if its asymmetry is similar to that of
  ordinary baryons. We show that such particles can solve the `solar
  composition problem'. The predicted small decrease in the low energy
  neutrino fluxes may be measurable by the Borexino and SNO+ experiments.
\end{abstract} 
\maketitle

We consider the capture by the Sun of asymmetric dark matter (ADM)
particles which have a relic asymmetry just as do baryons, in contrast
to the usual candidates for cold dark matter (CDM) such as
supersymmetric neutralinos which have a relic thermal abundance
determined by `freeze-out' from chemical equilibrium. Hence ADM does
not annihilate upon capture in astrophysical bodies such as the Sun,
leading to a build up of its concentration. In particular,
self-interactions can lead to an {\em exponential} increase of the ADM
abundance in the Sun as it orbits around the Galaxy, accreting dark
matter.

ADM does not have the usual indirect signatures e.g. there will be no
high energy neutrino signal from annihilations in the Sun. Instead ADM
will alter heat transport in the solar interior thus affecting the low
energy neutrino flux. This had been proposed as a solution to the
`solar neutrino problem'
\cite{Spergel:1984re,Faulkner:1985rm,Gilliland:1986}. Although the
solution is now understood to be neutrino oscillations
\cite{Bahcall:2004mz}, small changes induced by accreted CDM particles
may account for the current discrepancy \cite{PenaGaray:2008qe}
between helioseismological data and the revised `Standard Solar Model'
(SSM).

An asymmetry in ADM similar to that in baryons naturally explains why
their observed abundances are of the same order of magnitude. If ADM
arises from a strongly coupled theory (like the baryon of QCD), then
there is a conserved $U(1)$ global symmetry (like $B$ number in QCD)
which guarantees stability of the lightest $U(1)$ charged
object. Technicolour models of electroweak symmetry breaking
\cite{Weinberg:1979bn} provide an example of ADM from strong dynamics
in the form of the lightest neutral technibaryon
\cite{Nussinov:1985xr}. Recently, new viable types of technibaryon
dark matter (TIMPs) \cite{Gudnason:2006ug} as well as other particle
candidates for ADM \cite{Kaplan:1991ah} have been suggested.

While gravitational instability in collisionless CDM provides a good
explanation for the large-scale structures of the universe,
observations on galactic and smaller scales suggest that CDM may be
self-interacting \cite{Spergel:1999mh,Wandelt:2000ad}. If ADM arises
as a composite TIMP or a `dark baryon' from a strongly coupled model,
then it would naturally have such self-interactions.

We consider how the capture of self-interacting ADM by the Sun can
alter helioseismology and low energy neutrino fluxes.

\section{Capture of self-interacting ADM}

We refer to earlier discussions of the capture by the Sun of heavy
Dirac neutrinos having an asymmetry \cite{Griest:1986yu}, and of
symmetric CDM with self-interactions \cite{Zentner:2009is}. The
capture rate for CDM particles $\chi$ with {\em both} an asymmetry and
self-interactions is governed by the equation:
\begin{equation}
\label{Nddm}
\frac {\mathrm{d}N_\chi} {\mathrm{d}t} = C_\mathrm{\chi N} + C_{\chi\chi} N_\chi.
\end{equation}
Here $C_\mathrm{\chi \cal{N}}$ is the usual rate of capture of CDM
particles by scattering off nuclei (mainly protons) within the Sun,
while $C_{\chi\chi}$ is the rate of self-capture through scattering
off already captured $\chi$ particles. Hence the number of captured
particles would have grown as
\begin{equation}
\label{Nddm'}
N_\chi (t) = \frac{C_\mathrm{\chi \cal{N}}}{C_{\chi\chi}} 
\left(\mathrm{e}^{C_{\chi\chi} t} -1 \right),
\end{equation}
i.e. {\em exponentially} for $t \gtrsim C_{\chi\chi}^{-1}$. However
the effective cross-section for self-captures cannot increase beyond
$\pi r_\chi^2$ where $r_\chi$ is the scale-height of the region where
they are gravitationally trapped \cite{Spergel:1984re}. The linear
growth by contrast can continue up to the `black disk' limit i.e. $\pi
R_\odot^2$. In both cases there is an additional enhancement due to
`gravitational focussing' \cite{Spergel:1984re,Gould:1987ju} as we
quantify later.
The ejection of captured ADM particles by recoil effects in the
self-scattering can be neglected \cite{Zentner:2009is} and evaporation
is negligible for a mass exceeding 3.7 GeV \cite{Gould:1987ju}.

The ADM capture rate through spin-independent (SI) and spin-dependent
(SD) interactions can be written \cite{Jungman:1995df}:
\begin{equation} 
\label{eq-CaptureRate}
C_\mathrm{\chi \cal{N}}^\mathrm{SI,SD} = c_\mathrm{\chi \cal{N}}^\mathrm{SI,SD} 
\left(\frac{\rho_\mathrm{local}}{0.4~\mathrm{GeV\,cm}^{-3}}\right)
\sum_i \mathcal{F}_i \left(\frac{\sigma_i^\mathrm{SI,SD}}{10^{-40}\mathrm{cm}^2} 
\right)
\end{equation}
where $c_\mathrm{\chi \cal{N}}^\mathrm{SI} = 6.4 \times
10^{24}\mathrm{s}^{-1}$ and $c_\mathrm{\chi \cal{N}}^\mathrm{SD} = 1.7
\times 10^{25}\mathrm{s}^{-1}$, $\rho_\mathrm{local}$ is the estimated
local CDM density, and $\mathcal{F}_i (m_\chi)$ encodes the form
factors for different nuclei $i$ weighted by the solar chemical
composition --- the sum is over all nuclei (hydrogen only for SD
interactions). Here $\sigma_\mathrm{\chi \cal{N}}^\mathrm{SI,SD}$ is
the ADM-{\em nucleus} cross-section, which is related to
$\sigma_\mathrm{\chi N}$, the ADM-nucleon cross-section
\cite{Jungman:1995df}.
%
%
For spin-independent interactions, $\sigma_\mathrm{\chi N}$ is
constrained by direct detection experiments such as CDMS-II
\cite{Ahmed:2009zw}, XENON10 \cite{Angle:2009xb} and CoGeNT
\cite{Aalseth:2010vx} to be $\lesssim 10^{-39}~\mathrm{cm}^2$ for
$m_\chi = 5$~GeV. For spin-dependent interactions the constraints are
considerably weaker, e.g. PICASSO \cite{Archambault:2009sm} sets the
strongest bound of $\lesssim 10^{-36}~\mathrm{cm}^2$ for this mass.

Next we consider the self-capture rate in the Sun \cite{Zentner:2009is}:
\begin{equation}
\label{selfcapturerate}
C_{\chi\chi} = \sqrt{\frac{3}{2}}\ \rho_\mathrm{local}\ s_\chi \  
\frac{v_\mathrm{esc}^2(R_\odot)}{\bar{v}}\ \langle\phi\rangle\ 
\frac{\rm{erf(\eta)}}{\eta}
\end{equation}
where $s_\chi \equiv \sigma_{\chi\chi}/m_\chi$ is the ADM
self-interaction cross section divided by its mass, and
$v_\mathrm{esc} (R_\odot) \sim 618~\mathrm{km} \mathrm{s}^{-1}$ is the
escape velocity at the surface of the Sun, which is assumed to be
moving at $v_\odot = 220~\mathrm{km} \mathrm{s}^{-1}$ through a
Maxwell-Bolzmann distribution of CDM particles with velocity
dispersion, $\bar{v} \sim 270~\mathrm{km} \mathrm{s}^{-1}$. Here
$\langle\phi\rangle \sim 5.1$ is the average over $\phi (r) \equiv
v_\mathrm{esc}^2(r)/v_\mathrm{esc}^2(R)$ and $\eta \equiv
\sqrt{3/2}v_\odot/\bar{v}$.

Self-interacting CDM was proposed \cite{Spergel:1999mh} to account for
observations of galactic and subgalactic structure on scales
$\lesssim$ a few Mpc which are not in accord with numerical
simulations using collisionless cold particles. The discrepancy can be
solved if CDM has a mean free path against self-interactions of
$\lambda \sim 1~\mathrm{kpc} - 1~\mathrm{Mpc}$ corresponding to a
self-scattering cross-section between $s_\chi \sim 8 \times 10^{-22}$
and $\sim 8 \times 10^{-25}~\mathrm{cm}^2 \mathrm{GeV}^{-1}$
\cite{Spergel:1999mh}. A detailed analysis sets an upper limit of
$s_\chi \lesssim 10^{-23}~\mathrm{cm}^2 \mathrm{GeV}^{-1}$
\cite{Wandelt:2000ad}, while a study \cite{Randall:2007ph} of the
colliding `Bullet cluster' of galaxies implies a stronger bound of
$\sim 2 \times 10^{-24}~\mathrm{cm}^2 \mathrm{GeV}^{-1}$, which we
adopt for our calculations below.

A `dark baryon' from a QCD-like strongly interacting sector but with a
mass of about 5 GeV is a natural candidate for ADM.  Its relic density is
linked to the relic density of baryons via $\Omega_\chi \sim (m_\chi
{\cal N}_\chi/m_\mathrm{B} {\cal N}_\mathrm{B})\Omega_B$ where ${\cal
  N}_{\mathrm{B}, \chi}$ are the respective asymmetries. If ${\cal
  N}_\mathrm{B} \sim {\cal N}_\chi$ (e.g. if both asymmetries are
created by `leptogenesis' \cite{Davidson:2008bu}) then the required
CDM abundance is realised naturally. The self-interaction
cross-section of such a neutral particle can be estimated by scaling
up the neutron self-scattering cross-section $\sim 10^{-23}$~cm$^2$
\cite{Gardestig:2009zz} as: $\sigma_{\chi\chi} =
(m_\mathrm{n}/m_\chi)^2 \sigma_\mathrm{nn}$
which is just of the required order.  Note that the self-annihilation
cross-section will be of the same order which ensures that the ADM
thermal ({\em symmetric}) relic abundance is negligible, just as it is
for baryons.

Photon exchange, via a magnetic moment of the dark baryon, will give
rise to both spin-independent and spin-dependent interactions of
$\chi$ with nucleons. Recently this has been investigated in a model
of a 5 GeV dark baryon in a `hidden sector' interacting with the
photon through mixing with a hidden photon magnetic moment
\cite{An:2010kc}. From this model we infer that spin-independent
cross-section with nuclei of ${\cal O}(10^{-39})$~cm$^2$ can be
achieved. Moreover this will be accompanied by spin-dependent
interactions which would aid further in the heat transport in the Sun
as discussed below.  Since the photon couples {\em only} to the proton
in direct detection experiments, the limit on $\sigma_\mathrm{\chi N}$ is
degraded for this model to $\sim 4 \times 10^{-39}$~cm$^2$ which we
adopt as an example later.

\section{Helioseismology and Solar neutrinos}

\begin{figure}[htp!]
{\includegraphics[width=\linewidth]{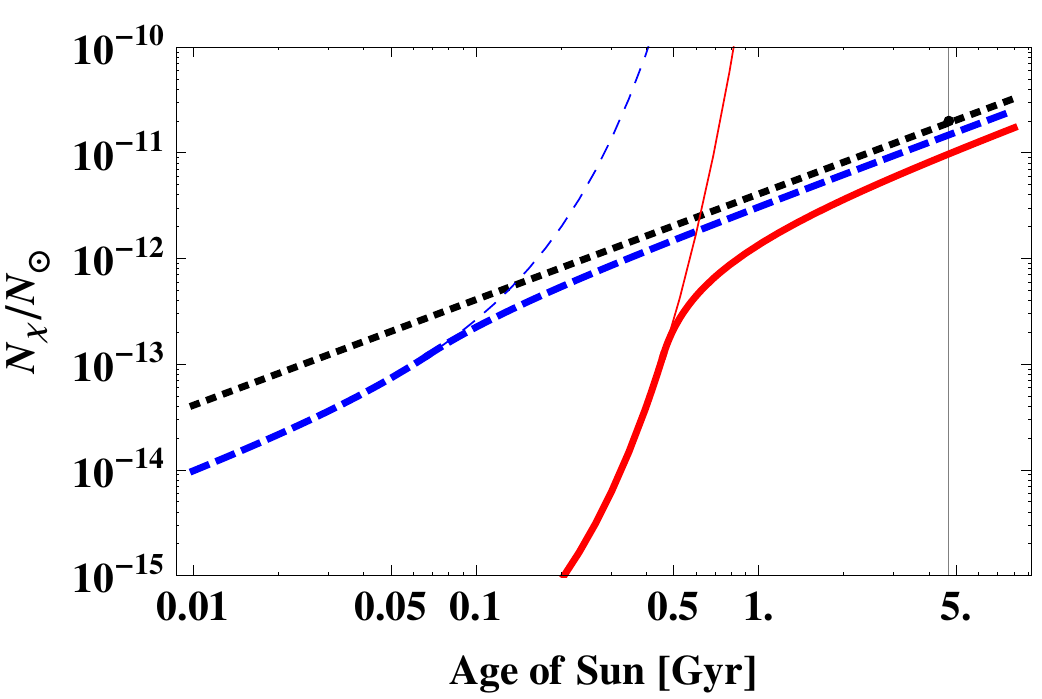}}
\caption{Growth of the relative abundance of 5 GeV mass ADM particles
  in the Sun until its present age (vertical line) assuming
  $s_{\chi\chi} = 2 \times 10^{-24} \mathrm{cm}^2 \mathrm{GeV}^{-1}$,
  and $\sigma_{\chi N} = 10^{-39}~\mathrm{cm}^2$ (solid line) and
  $10^{-36}~\mathrm{cm}^2$ (dashed line), these being the maximum
  experimentally allowed values for spin-independent and
  spin-dependent interactions respectively. Also shown is the `black
  disk' limit (dotted line) for the Sun.}
\label{Ncaptmax}
\end{figure}

Fig.~\ref{Ncaptmax} shows the growth of the number of captured ADM
particles in ratio to the number of baryons in the Sun, for a
scattering cross-section on nucleons as large as is
experimentally allowed, including the `gravitational focussing' factor
of $(v_\mathrm{esc}(r)/\bar{v})^2$ \cite{Gould:1987ju} and
setting $r=R_\odot$ or $r_\chi$ ($\simeq 0.07 R_\odot$ for $m_\chi =
5$~GeV) as appropriate.

Note that due to the self-captures, the limiting abundance $N_\chi/N_\odot \sim
2 \times 10^{-11}$ is almost independent of the actual scattering
cross-section. Such an ADM fraction in the Sun can affect the thermal
conductivity and thereby solar neutrino fluxes
\cite{Faulkner:1985rm,Spergel:1984re}. The SSM \cite{Bahcall:2004pz}
predicts 3 times the observed neutrino flux (the `Solar neutrino
problem') but this is now well understood taking into account neutrino
oscillations \cite{Bahcall:2004mz}. Moreover until recently, the SSM
with the `standard' solar composition \cite{Grevesse:1998bj} agreed
very well with helioseismology \cite{Serenelli:2009ww}. However the
revision of the solar composition \cite{Asplund:2009fu} means that the
SSM no longer reproduces the sound speed and density profile so there
is now a `solar composition problem' \cite{PenaGaray:2008qe}. We show
that the presence of ADM in the Sun can resolve this problem and
precision measurements of solar neutrino fluxes can constrain the
properties of self-interacting ADM.

A simple scaling argument gives for the luminosity carried by the
ADM \cite{Spergel:1984re}:
\begin{equation}
  L_\chi \sim 4 \times 10^{12} L_{\odot} \frac{N_\chi}{N_\mathrm{\odot}}
  \frac{\sigma_\mathrm{\chi N}}{\sigma_\odot}\sqrt{\frac{m_\mathrm{N}}{m_\chi}}\ , 
\label{lumisimple}
\end{equation}
where $L_{\odot}\sim 4 \times 10^{33} \rm{ergs}\, \rm{s}^{-1}$. When
the ADM mean free path $\lambda_\chi$ is large compared to the
scale-height $r_\chi$
then the energy transfer is {\em non-local}. This is the case when
$\sigma_\mathrm{\chi N} \ll \sigma_{\rm{\odot}}$ where $\sigma_\odot
\equiv (m_\mathrm{N}/M_\odot) R_\odot^2 \sim 4 \times
10^{-36}~\mathrm{cm}^2$ is a critical scattering cross-section. We
consider the ADM trapped in the Sun as an isothermal gas at
temperature $T_\chi$ \cite{Spergel:1984re}, so the luminosity
$L_\mathrm{\chi}$ carried by the particles is:
\begin{equation}
  L_\mathrm{\chi}(r) = \int^{r}_{0} \mathrm{d}r' \; 
   4 \pi r'\,^2\,\rho(r') \,\epsilon_\chi(r') \ , 
\end{equation}
where $\epsilon_\chi (r') \propto (T(r') - T_{\chi} ) N_\chi
\sigma_\mathrm{\chi N}$ is the energy transferred to the ADM per
second per gram of nuclear matter and $\rho(r')$ is the density in the
Sun \cite{Spergel:1984re}.
%
%
The ADM temperature $T_{\chi}$ is fixed by requiring that the energy
absorbed in the inner region ($T(r) > T_{\chi}$) is equal to that
released in the outer region ($T(r) < T_{\chi}$), such that $L_{\rm
  \chi}(R_{\odot}) = 0$.

\begin{figure}[htp!]
{\includegraphics[width=\linewidth]{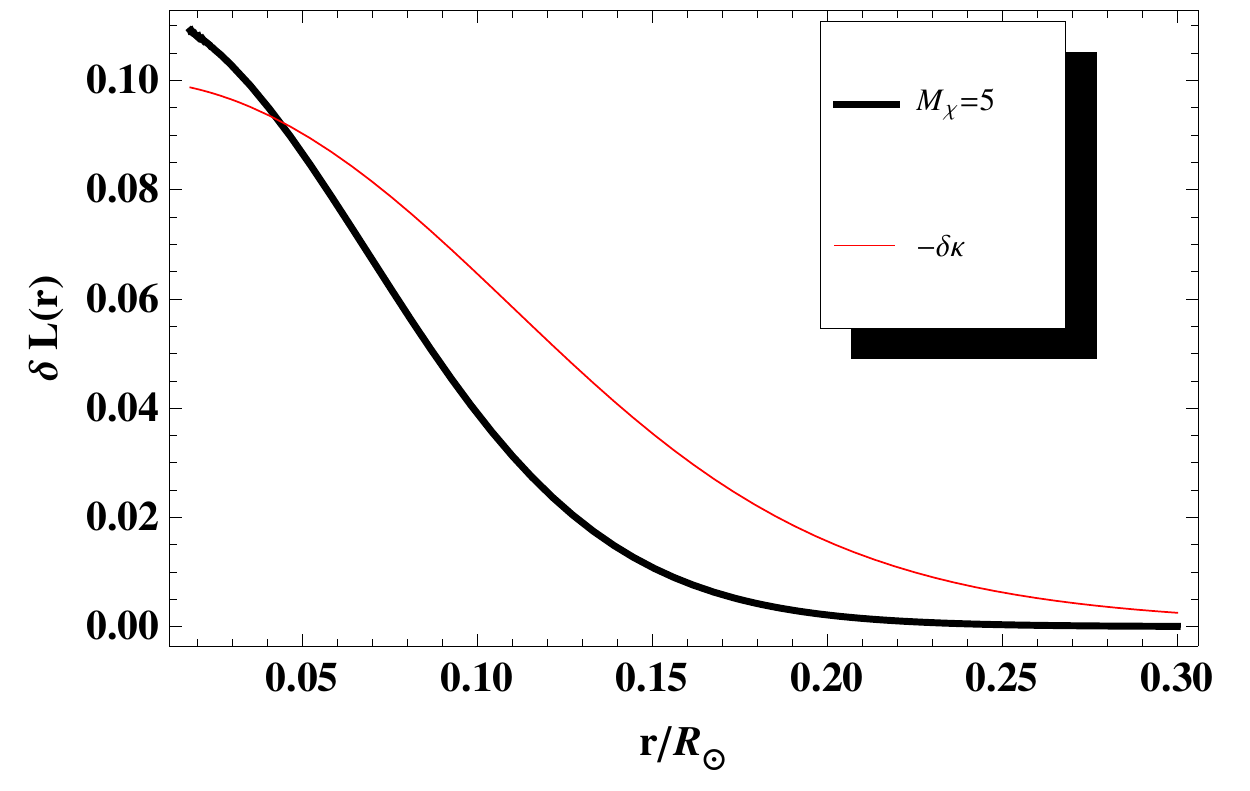}}
\caption{The radial variation of $\delta L(r) \equiv
  L_\chi(r)/L_{\odot}(r)$ due to ADM of mass 5 GeV, using the
  approximation of Ref.\cite{Spergel:1984re} and $L_\odot (r)$ from
  the BS05 (OP) Standard Solar Model \cite{Bahcall:2004pz}.}
\label{luminosityfunction}
\end{figure}

This approximation overestimates the energy transfer by a small factor
\cite{Gould:1989ez,Dearborn:1990mm} but is sufficiently accurate for
the present study. We adopt a simple polytropic model for the Sun's
temperature $T$, number density $n_\mathrm{p}$ and gravitational
potential $V$ \cite{Spergel:1984re}. The resulting variation of the
solar luminosity $\delta L(r) \equiv L_\chi(r)/L_{\odot}(r)$ is shown
in Fig.\ref{luminosityfunction} assuming $\sigma_\mathrm{\chi N} = 4
\times 10^{-39}~\mathrm{cm}^2$ (i.e. $10^{-3} \sigma_\mathrm{\odot}$)
and $N_\chi = 2 \times 10^{-11} N_\mathrm{\odot}$ from
Fig~\ref{Ncaptmax}. Note that the luminosity scales linearly with both
$\sigma_{\mathrm{\chi N}}$ and $N_\chi/N_\odot$.

From the radiative transport equation it follows that a small
variation of the solar luminosity is equivalent to an {\em opposite}
small variation in the effective radiative opacity: $\delta L(r) \sim
-\delta\kappa_\gamma(r) \equiv -\kappa_\chi(r)/\kappa_\gamma(r)$
\cite{Bottino:2002pd}. The effect of such a localised opacity
variation in the region $r \lesssim 0.2 R_\odot$ has been studied by a
Monte Carlo simulation \cite{Fiorentini:2001et} and results in
excellent agreement obtained using a linear approximation to the solar
structure equations
\cite{Villante:2009xs}. Fig.~\ref{luminosityfunction} shows that the
opacity modification due to a 5 GeV ADM with a relative concentration
of $10^{-11}$ is roughly equivalent to the effect of a $~10\%$ opacity
variation. The impact of this luminosity variation on neutrino fluxes
can be estimated by evaluating $\delta L(r)$ at the scale height of
the ADM distribution, $\delta L (r_\chi)$ \cite{Spergel:1999mh}. In
general, to have an observable effect requires $\sigma_\mathrm{\chi N}
N_\chi / \sigma_\odot N_\odot \gtrsim 10^{-14}$.

It is possible through helioseismology to determine e.g. the mean
variations of the sound speed profile $\langle \delta c/c \rangle$ and
density $\langle\delta\rho/\rho\rangle$ of the Sun, as well as the
boundary of the convective zone $R_\mathrm{CZ}$. In particular
$R_\mathrm{CZ}$ is determined to be $(0.713 \pm 0.001) R_\odot$ while
the SSM with the revised composition \cite{Asplund:2009fu} predicts
values that are too high by up to 15$\sigma$
\cite{Serenelli:2009ww}. Lowering the opacity in the central region of
the Sun with ADM also lowers the convective boundary. The $~10\%$
opacity variation shown in Fig.~\ref{luminosityfunction} leads to a
$\sim 0.7\%$ reduction in $R_\mathrm{CZ}$ \cite{Villante:2009xs} and
thus {\em restores} the agreement with helioseismology. The sound
speed and density profiles which are presently underestimated in the
region $0.2 R_\odot \lesssim r \lesssim R_\mathrm{CZ}$ would also be
corrected by the opacity modification displayed in
Fig.~\ref{luminosityfunction}.

\begin{figure}[htp!]
{\includegraphics[width=\linewidth]{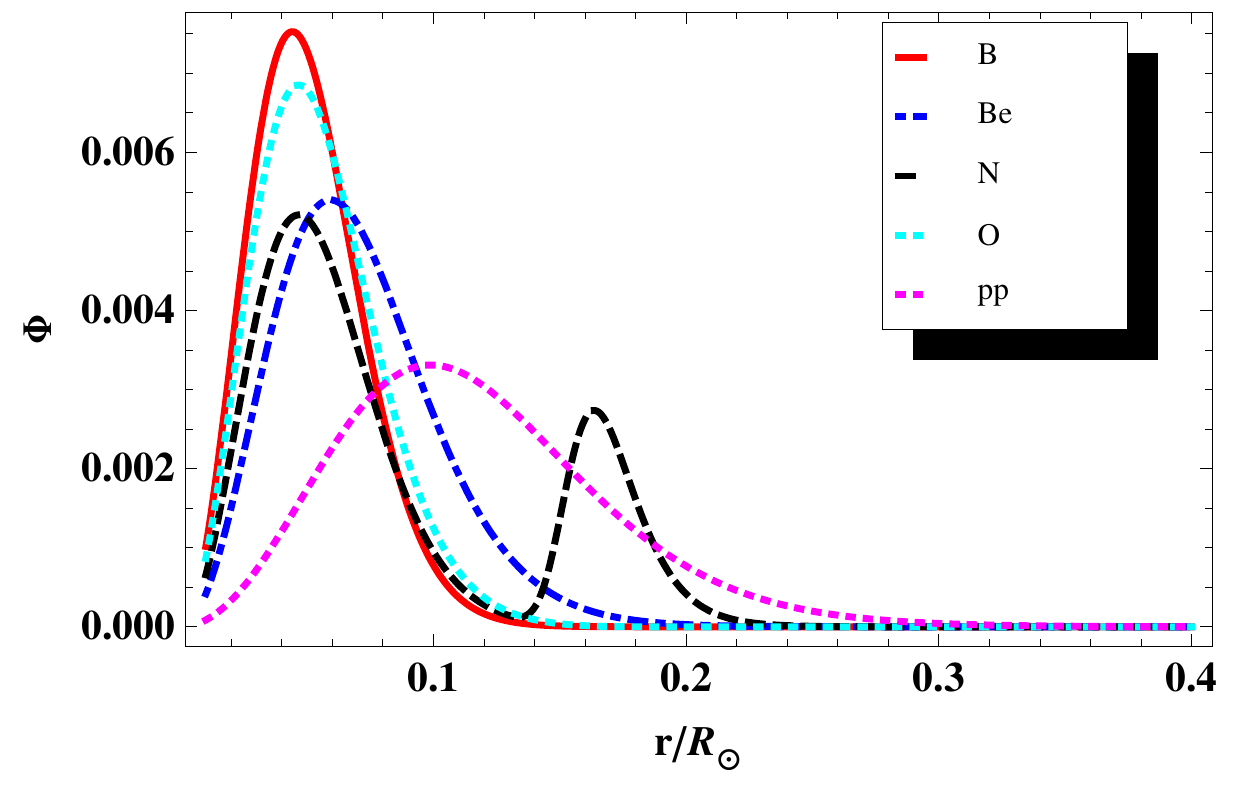}}
\caption{Neutrino producing regions in the solar
  interior}\label{neutrinosBPS05}
\end{figure}
The modification of the luminosity profile extends into the neutrino
producing region as displayed for the SSM in
Fig.~\ref{neutrinosBPS05}. Comparing with
Fig.~\ref{luminosityfunction} we see that precision measurements of
different neutrino fluxes may be able to test the ADM model and
determine its parameters. The ADM mass determines the scale height
$r_\chi$, hence the relative modifications of individual neutrino
fluxes, while the cross-section determines the capture rate and
thereby the overall modification. Both Monte Carlo simulations
\cite{Fiorentini:2001et} and the `linear solar model'
\cite{Villante:2009xs} show that the variation of neutrino fluxes with
respect to localised opacity changes in the neutrino producing region
($r \lesssim 0.2 R_\odot$) scales approximately as $\delta
\Phi_\mathrm{B} \sim 1.5 \delta\kappa$ and $\delta \Phi_\mathrm{Be}
\sim 0.7 \delta\kappa$. The opacity variation in
Fig.~\ref{luminosityfunction} leads to variations
$\delta\Phi_\mathrm{B} = -17\% , \delta\Phi_\mathrm{Be} = -6.7\%$ and
$\delta\Phi_\mathrm{N} = -10\% , \delta\Phi_\mathrm{O} = -14\%$
\cite{Villante:2009xs}. Measurements of the $^8$B flux by
Super-Kamiokande \cite{:2008zn}, SNO \cite{Aharmim:2008kc} and
Borexino \cite{Collaboration:2008mr} are precise to 10\% while the
expectations vary by up to 20\% depending on whether the old
\cite{Grevesse:1998bj} or the new \cite{Asplund:2009fu} composition is
used \cite{Serenelli:2009ww}. For the $^7$Be flux, the theoretical
uncertainty is 10\%, while Borexino aims to make a measurement precise
to 3\% \cite{Arpesella:2008mt}. SNO+ is expected to make a first
measurement of the pep and CN-cycle fluxes \cite{Haxton:2008yv}. Thus
the effects of metallicity and luminosity variations can be
distinguished in principle.

\section{Conclusions}

Asymmetric dark matter does not annihilate upon capture in the Sun and
can therefore affect heat transport in the solar interior and
consequently neutrino fluxes. This is particularly true for particles
with self-interactions which would also explain the paucity of
sub-galactic structure. We have shown that the presence of such
particles in the Sun can solve the `solar composition problem'.

Intriguingly a 5 GeV `dark baryon' would naturally a) have the
required relic abundance if it has an initial asymmetry similar to
that of baryons, b) have a self-interaction cross-section of the right
order to explain sub-galactic structure, c) modify the deep interior
of the Sun, restoring agreement between the standard solar model and
helioseismology, and d) be consistent with recent hints of signals in
direct detection experiments \cite{Ahmed:2009zw,Aalseth:2010vx}. Such
a 5 GeV ADM particle would lower the solar neutrino fluxes which ought
to be testable by the Borexino and (forthcoming) SNO+ experiments.


{\bf Note added:} After this paper was submitted to arXiv (1003.4505),
another study appeared \cite{Cumberbatch:2010hh} with similar findings
concerning the effect of ADM on helioseismology and neutrino fluxes. However
a second such study \cite{Taoso:} finds negligible effects using a solar model simulation.

\section*{Acknowledgements}

We thank G. Bertone, C. Kouvaris, F. Sannino, J. Silk and F. Villante
for correspondence, and especially S. Nussinov for pointing out an
error in an earlier version. MTF acknowledges a VKR Foundation
Fellowship. SS acknowledges support by the EU Marie Curie Network ``UniverseNet''
(HPRN-CT-2006-035863).

\end{document}